\documentclass{llncs}

\usepackage[all]{xy}
\usepackage{comment}
\usepackage{amsfonts,amsmath,latexsym,amssymb}
\usepackage{tikz}
\usepackage{listings}
\usepackage{color}
\usepackage{algpseudocode}
\usepackage{algorithm}

\newcommand{\Nat}{{\mathbb N}}

\newcommand{\kw}[1]{\ensuremath{\mathsf{#1}}}

\newcommand{\cat}[1]{\ensuremath{\mathbf{#1}}}

\begin{document}
\title{A Productivity Checker for Logic Programming}

\author{E. Komendantskaya\inst{1} and P. Johann\inst{2} and M.Schmidt\inst{3}}
\institute{Heriot-Watt University, Edinburgh, Scotland, UK \and Appalachian State University, Boone, NC, USA \and University of Osnabr\"uck, Osnabr\"uck, Germany }

\maketitle

\begin{abstract}
Automated analysis of recursive derivations in logic programming is
known to be a hard problem.  Both termination and non-termination are
undecidable problems in Turing-complete languages.  However, some
declarative languages offer a practical work-around for this problem,
by making a clear distinction between whether a program is meant to be
understood inductively or coinductively. For programs meant to be
understood inductively, termination must be guaranteed, whereas for
programs meant to be understood coinductively,
productive non-termination (or ``productivity") must be ensured.  In
practice, such classification helps to better understand and implement
some non-terminating computations.

Logic programming was one of the first declarative languages to make
this distinction: in the 1980's, Lloyd and van Emden's ``computations
at infinity" captured the big-step operational semantics of
derivations that produce infinite terms as answers. In modern terms,
computations at infinity describe ``global productivity" of
computations in logic programming. Most programming languages
featuring coinduction also provide an observational, or small-step,
notion of productivity as a computational counterpart to global
productivity. This kind of productivity 
is ensured by checking that finite initial fragments of infinite
computations can always be observed to produce finite portions of
their infinite answer terms.

In this paper we introduce a notion of {\em observational
  productivity} for logic programming as an algorithmic approximation
of global productivity, give an effective procedure for semi-deciding
observational productivity, and offer an implemented automated
observational productivity checker for logic programs.
\end{abstract}
\vspace*{-0.1in}



\vspace*{-0.2in}

\section{Introduction}\label{sec:intro}

Induction is pervasive in programming and program verification. It
arises in definitions of finite data (e.g., lists, trees, and other
algebraic data types), in program semantics (e.g., of finite iteration
and recursion), and proofs (e.g., of properties of finite data and
processes). Coinduction, too, is important in these arenas, arising in
definitions of infinite data (e.g., lazily defined infinite streams),
in program semantics (e.g., of concurrency), and in proofs (e.g., of
observational equivalence, or bisimulation, of potentially infinite
processes). It is thus desirable to have good support for both
induction and coinduction in systems for reasoning about programs.

Given a logic program $P$ and a term $A$, SLD-resolution provides a
mechanism for automatically (and inductively) inferring that $P \vdash
A$ holds, i.e., that $P$ logically entails $A$.  The ``answer'' for a
program $P$ and a query $? \gets A$ is a substitution $\sigma$
computed from $P$ and $A$ by SLD-resolution. Soundness of
SLD-resolution ensures that $P \vdash \sigma(A)$ holds, so we also say
that $P$ \emph{computes} $\sigma(A)$.
\begin{example}[Inductive logic program]\label{ex:der} 
The program $P_1$ codes the Peano numbers:

\vspace*{0.05in}\noindent
0. $\mathtt{nat(0)} \; \gets \;$\\ 
1. $\mathtt{nat(s(X))} \; \gets \; \mathtt{nat(X)}$

\vspace*{0.05in}\noindent To answer the question ``{\em Does $P_1
  \vdash \mathtt{nat(s(X))}$ hold?}'', we represent it as the logic
programming (LP) query $?  \gets \mathtt{nat(s(X))}$ and resolve it 
with $P_1$. It is standard in implementations of traditional LP to use
a topmost clause selection strategy, which resolves goals against
clauses in the order in which they appear in the program. Topmost
clause selection gives the derivation
$\mathtt{nat(s(X))} \rightarrow \mathtt{nat(X)} \rightarrow
\mathtt{true}$ for $P_1$ and $? \gets \mathtt{nat(s(X))}$, which computes 
the answer $\{\mathtt{X \mapsto 0}\}$ in its last step. Since $P_1$
computes $\mathtt{nat(s(0))}$, one answer to our question is ``Yes,
provided $\mathtt{X}$ is $\mathtt{0}$.''
\end{example}

While inductive properties of terminating computations are quite well
understood~\cite{Llo88}, non-terminating LP computations
are notoriously difficult to reason about, and can arise even for
programs that are intended to be inductive:
\begin{example}[Coinductive meaning of inductive logic program]\label{ex:ordinal}
If $P_1'$ is obtained by reversing the order of the clauses in the
program $P_1$ from Example~\ref{ex:der}, then the SLD-derivation for
program $P_1'$ and query $? \gets \mathtt{nat(s(X))}$ does not
terminate under standard topmost clause selection. Instead, it results
in an attempt to compute the ``answer'' $\{\mathtt{X}
\mapsto \mathtt{s(s(...))}\}$ by repeatedly resolving with Clause
1. Nevertheless, $P_1'$ is still computationally meaningful, since it
computes the first limit ordinal at infinity~\cite{Llo88}.
\end{example}
\noindent
Some programs do not admit terminating computations under
{\em any} 
selection strategy:
\begin{example}[Coinductive logic program]\label{ex:zstream} 
No derivation for the query $? \gets \mathtt{stream(X)}$ and the
program $P_2$ comprising the clause

\vspace*{0.05in}\noindent
0. $\mathtt{stream(scons(0,Y))} \; \gets \;
\mathtt{stream(Y)}$

\vspace*{0.05in}\noindent terminates with an answer, be it success or
otherwise. Nevertheless, $P_2$ has computational meaning:
it computes the infinite stream of $0$s
at infinity.
\end{example}

The importance of developing sufficient infrastructure to support
coinduction in automated proving has been argued across several
communities; see, e.g.,~\cite{LeinoM14,ReynoldsB15,SimonBMG07}.  In
LP, the ability to work with non-terminating and coinductive programs
depends crucially on understanding the structural properties of
non-terminating SLD-derivations. To illustrate, consider the
non-terminating programs $P_3$, $P_4$, and $P_5$:

\noindent
\begin{tabular}{p{2cm} p{4cm} p{7cm}}
\hline
~~Program & Program definition & For query $? \gets \mathtt{p(X)}$,
computes the answer:\\ ~~~~$P_3$ & ~~$\mathtt{p(X) \gets p(X)}$ &
~~$\textit{id}$ \\ ~~~~$P_4$ & ~~$\mathtt{p(X) \gets p(f (X))}$ &
~~$\textit{id}$ \\ ~~~~$P_5$ & ~~$\mathtt{p(f ( X)) \gets p(X)}$ &
~~$\{\mathtt{X} \mapsto \mathtt{f (f ...)}\}$ \\
\hline
\end{tabular}

\noindent
Programs $P_3$ and $P_4$ each loop without producing any substitutions
at all; only $P_5$ computes an infinite term at infinity. It is of
course not a coincidence that only $P_5$ resembles a (co)inductive
data definition by pattern matching on a constructor, as is commonly
used in functional programming.

When an infinite SLD-derivation computes an infinite object, and this
object can be successively approximated by applying to the initial
query the substitutions computed at each step of the derivation,
the derivation is said to be {\em globally productive}.  
The only derivation for program $P_5$
and the query $? \gets \mathtt{p(X)}$ is globally productive since it
approximates, in the sense just described, the infinite term
$\mathtt{p(f(f ... ))}$.  In terminology of~\cite{Llo88}, it computes $\mathtt{p(f(f ... ))}$ at infinity.  Programs $P_2$ and $P'_1$ similarly give
rise to globally productive derivations. But no derivations for $P_3$
or $P_4$ are globally productive.

Since global productivity determines which non-terminating logic
programs can be seen as defining coinductive data structures, we would
like to identify exactly when a program is globally productive.
But porting functional programming methods of ensuring productivity by
static syntactic checks is hardly possible.  Unlike pattern matching
in functional programming, SLD-resolution is based on {\em
unification}, which has very different operational properties ---
including termination and productivity properties --- from pattern
matching.  For example, programs $P_1$, $P_1'$, $P_2$, and $P_5$ are
all terminating by term-matching SLD-resolution, i.e., resolution in which unifiers are restricted to matchers, as in
term rewriting. We thus call this kind of derivations \emph{rewriting derivations}.

\vspace{-0.1in}
\begin{example}[Coinductive program defining an irrational infinite term]\label{ex:from-a}
The program $P_6$ comprises the single clause

\vspace*{0.05in}\noindent
0. $\mathtt{from(X, scons(X,Y))} \gets \mathtt{from(s(X),Y)}$

\vspace*{0.05in}\noindent
For $P_6$ and the query $? \gets
\mathtt{from(0,Y)}$, SLD-resolution computes at infinity the answer substitution
$\{\mathtt{Y} \mapsto [\mathtt{0, s(0)}$,
  $\mathtt{s(s(0)),\ldots}]\}$. Here $[t_1, t_2, \ldots]$ abbreviates
  $\mathtt{scons}(t_1, \mathtt{scons}(t_2, \ldots))$, and similarly
  in the remainder of this paper.  This derivation depends crucially on
  unification since variables occurring in the two arguments to
  $\mathtt{from}$ in the clause head overlap. If we restrict to
  rewriting, then there are no successful derivations (terminating or
  non-terminating) for this choice of program and query.
\end{example}
\vspace{-0.1in}

Example~\ref{ex:from-a} shows that any analysis of global productivity
must necessarily rely on specific properties of the operational
semantics of LP, rather than on program syntax alone. It has been
observed in~\cite{KPS12-2,JKK15} that one way to distinguish
globally productive programs operationally is to identify those that
admit infinite SLD-derivations, but for which rewriting derivations
always terminate.  We call this program property 
\emph{observational productivity}. The programs $P_1$, $P_1'$, $P_2$,
$P_5$, $P_6$ are all observationally productive.

The key observation underlying observational productivity is that
terminating rewriting derivations can be viewed as points of finite
observation in infinite derivations. Consider again program $P_6$ and
query $?  \gets \mathtt{from(0,Y)}$ from
Example~\ref{ex:from-a}. Drawing rewriting derivations vertically and
unification-based resolution steps horizontally, we see that each
unification substitution applied to the original query effectively
observes a further fragment of the stream computed at infinity:

\vspace*{0.1in}
\hspace{-0.1in}
\begin{tikzpicture}[scale=0.30,baseline=(current bounding box.north),grow=down,level distance=20mm,sibling distance=50mm,font=\footnotesize]
  \node {$\mathtt{from(0,X)}$};
  \end{tikzpicture}
$\stackrel{\{\mathtt{X} \mapsto \mathtt{[0,X']}\}}{\rightarrow}$\hspace*{0.2in}
\begin{tikzpicture}[scale=0.30,baseline=(current bounding box.north),grow=down,level distance=20mm,sibling distance=60mm,font=\footnotesize ]
  \node { $\mathtt{from(0,[0, X'])}$}
          child { node {$\mathtt{from(s(0),X')}$}};
  \end{tikzpicture}
	$\stackrel{\{\mathtt{X'}\mapsto \mathtt{[s(0),X'']}\}}{\rightarrow}$\hspace*{0.1in}
	\begin{tikzpicture}[scale=0.30,baseline=(current bounding box.north),grow=down,level distance=20mm,sibling distance=60mm,font=\footnotesize ]
  \node { $\mathtt{from(0,[0, s(0), X''])}$}
	child { node{ $\mathtt{from(s(0),[s(0),X''])}$ }
          child { node {$\mathtt{from(s(s(0)),X'')}$}}};
  \end{tikzpicture}~
$\stackrel{.}{\rightarrow} \ldots$
\vspace*{0.1in}

\noindent
If we compute unifiers only when rewriting derivations terminate,
then the resulting derivations exhibit consumer-producer behaviour:
rewriting steps consume structure (here, the constructor
$\mathtt{scons}$), and unification steps produce more structure (here,
new $\mathtt{scons}$es) for subsequent rewriting steps to
consume. This style of interleaving matching and unification steps was
called \emph{structural resolution} (or S-resolution)
in~\cite{JKK15,KJ15}.

Model-theoretic properties of S-resolution relative to least and
greatest Herbrand models of programs were studied in~\cite{KJ15}. In
this paper, we provide a suitable algorithm for semi-deciding
observational productivity of logic programs, and present its
implementation~\cite{Martin}, see also Appendix B online.  As exemplified above, observational
productivity of a program $P$ is in fact a conjunction of two
properties of $P$:

\vspace*{-0.07in}

\begin{enumerate}
\item \emph{universal observability}: termination of \emph{all}
rewriting derivations, and 
\item \emph{existential liveness}: existence of {\em at least one}
non-terminating S-resolution or SLD-resolution derivation.
\end{enumerate}

\vspace*{-0.07in}

\noindent
While the former property is universal, the latter must be
existential. For example, the program $P_1$ defining the Peano numbers
can have both inductive and coinductive meaning. When determining that
a program is observationally productive, we must certify that the
program actually \emph{does} admit derivations that produce infinite
data, i.e., that it actually {\em can} be seen as a coinductive
definition.  Our algorithm for semi-deciding observational
productivity therefore combines two checks:

\vspace*{-0.07in}

\begin{enumerate}
\item \emph{guardedness checks} that semi-decide
universal observability: if a program is guarded, then it is
universally observable. (The converse is not true in general.)
\item \emph{liveness invariant checks} ensuring that, if a program
is guarded and exhibits an invariant in its consumption-production of
constructors, then it is existentially live.
\end{enumerate}

\vspace*{-0.07in}

This is the first work to develop productivity checks for LP.  An
alternative approach to coinduction in LP, known as
CoLP~\cite{GuptaBMSM07,SimonBMG07}, detects loops in derivations and
closes them coinductively. However, loop detection was not intended as
a tool for the study of productivity and, indeed, is insufficient for
that purpose: programs $P_3$, $P_4$ and $P_5$, of which only the
latter is productive, are all treated similarly by CoLP, and all give
coinductive proofs via its loop detection mechanism.

Our approach also differs from the usual termination checking
algorithms in term-rewriting systems (TRS)~\cite{Terese,Arts2000,HM04}  and
LP~\cite{deSchreye1994199,Pf92,RohwedderP96,Schneider-KampGST06,NguyenGSS07}. Indeed, these algorithms
focus on guaranteeing termination, rather than productivity, see Section~\ref{sec:trs}. And
although the notion of  productivity has been studied
in TRS~\cite{EndrullisGHIK10,EndrullisHHP015}, the actual technical
analysis of productivity is rather different there because it
considers infinitary properties of rewriting, whereas observational productivity relies on termination of rewriting.

The rest of this paper is organised as follows. In
Section~\ref{sec:terms} we introduce a \emph{contraction ordering} on
terms that extends the more common lexicographic ordering, and argue
that this extension is needed for our productivity analysis. We also
recall that static guardedness checks do not work for LP. In
Section~\ref{sec:cotrees} we employ contraction orderings in dynamic
guardedness checks and present a decidable property, called $GC2$,
that characterises guardedness of a single rewriting derivation, and
thus certifies existential observability. In
Section~\ref{sec:guardedness} we employ $GC2$ to develop an algorithm,
called $GC3$, that analyses \emph{consumer-producer} invariants of
 S-resolution
derivations to certify universal observability. For universally
observable programs, these invariants also serve as liveness invariant
checks.  We also prove that $GC3$ indeed semi-decides observational
productivity. In Section~\ref{sec:trs} we discuss related work and in Section~\ref{sec:app} -- implementation and applications of the productivity checker. In Section~\ref{sec:conclusion} we conclude the
paper.

\vspace*{-0.19in}

\section{Contraction Orderings on Terms}\label{sec:terms} 

In this section, we will introduce the contraction ordering on first-order terms, on which our productivity checks will rely.
We work with the standard definition of first-order logic programs. 
 A \emph{signature} $\Sigma$ consists of a set  $\mathcal{F}$ of function
   symbols $f,g, \ldots$ each equipped with an arity.  Nullary (0-ary) function symbols are
 constants. We also assume a countable set $\mathit{Var}$ of variables, and a set $\mathcal{P}$ of predicate symbols each equipped with an arity.
We have the following standard definition for terms, formulae and Horn clauses:

\vspace*{-0.07in}
\begin{definition}[Syntax of Horn clauses and programs]\label{df:syntax}

Terms $Term \ ::= \ Var \ | \ \mathcal{F}(Term,..., Term)$

   Atomic formulae (or atoms) $At \ ::= \ \mathcal{P}(Term,...,Term)$

  (Horn) clauses $CH \ ::= \ At \gets At,..., At$
	
	Logic programs $Prog \ ::= CH, ... , CH$
\end{definition}
\vspace*{-0.07in}

In what follows, we will use letters $A,B$ with subscripts to refer to elements of $At$. 
Given a program $P$, we assume all clauses are indexed by natural numbers starting from $0$. When we need to refer to $i$th clause of program $P$, we will use
notation $P(i)$. To refer to the head of clause $P(i)$, we will use notation $\mathit{head}(P(i))$.



A \emph{substitution}  is a total function $\sigma:
\mathit{Var} \to Term$. 
Substitutions are extended from variables
to terms as usual: if $t\in Term$ and $\sigma$ is a substitution, then the {\em application}
$\sigma(t)$ is a result of applying $\sigma$ to all variables in $t$.
A substitution $\sigma$ is a \emph{unifier}
for $t, u$ if $\sigma(t) = \sigma(u)$, and is a
\emph{matcher} for $t$ against $u$ if $\sigma(t) = u$. 
A substitution $\sigma$ is a {\em most general unifier} ({\em mgu}) for
$t$ and $u$ if it is a unifier for $t$ and
$u$ and is more general than any other such unifier. A {\em most
  general matcher} ({\em mgm}) $\sigma$ for $t$ against $u$ is defined analogously. 

We can view every term and atom as a tree. Following standard definitions~\cite{Courcelle83,Llo88}, such trees can be indexed by
elements of a suitably defined tree language.
Let $\Nat^*$ be the set of all
finite words (i.e., sequences) over the set $\Nat$ of natural
numbers. A set $L \subseteq \Nat^*$ is a \emph{(finitely branching) tree
  language} if the following two conditions hold: (i) for all $w \in \Nat^*$ and all $i,j \in \Nat$, if $wj
\in L$ then $w \in L$ and, for all $i<j$, $wi \in L$, and 
(ii) for all
$w \in L$, the set of all $i\in \Nat$ such that $wi\in L$ is finite. A
tree language $L$ is {\em finite} if it is a finite subset of
$\Nat^*$, and {\em infinite} otherwise.
Term trees (for terms and atoms) are defined as mappings from a tree language $L$ to the given signature, see~\cite{Courcelle83,Llo88,JKK15}.
 Informally speaking,  every symbol occurring in a term or an atom 
receives an index from $L$. 

In what follows, we will work with term tree representation of all terms and atoms, and for brevity we will refer to all term trees simply as
 \emph{terms}. 
We will use notation $t(w)$ when we need to talk about the element of the term tree $t$ indexed by a word $w \in L$. Note that leaf nodes are always given by variables or constants.

\begin{example}
Given $L = \{ \epsilon, 0, 00, 01 \}$, the atom $\mathtt{stream(scons(0,Y))}$ can be seen as a term tree $t$ given by the map $t(\epsilon) = \mathtt{stream}$, 
$t(0) = \mathtt{scons}$, $t(00) = \mathtt{0}$, $t(01) = \mathtt{Y}$.
\end{example}
We can use such indexing to refer to subterms, and notation $\mathit{subterm}(t,w)$ will refer to a subterm of term $t$ starting at node $w$. In the above example,
where $t = \mathtt{stream(scons(0,Y))}$, 
$\mathit{subterm}(t,0)$ is $\mathtt{scons(0,Y)}$. 


Two most popular tools for termination analysis of
declarative programs are lexicographic ordering and (recursive) path ordering of terms. Informally,
the idea can be adopted to LP setting as follows. Suppose we have a
clause $A \gets B_1 , \ldots, B_i, \ldots , B_n$.  We may want to
check whether  each $B_i$ sharing the predicate with $A$  is
 ``smaller"' than $A$, since
this guarantees that no infinite rewriting derivation is triggered by
this clause. For lexicographic ordering we will write $B_i <_l A$ and for path ordering we will write $B_i <_p A$.

Using standard orderings to prove universal observability works well
for program $P_2$, 
since
$\mathtt{stream(Y)} $ $<_l \mathtt{stream(scons(0,Y))}$ and
$\mathtt{stream(Y)} <_p $ $ \mathtt{stream(scons(0,Y))}$, 
and so any
rewriting derivation for $P_2$ terminates. But universal observability
of $P_6$ from Example~\ref{ex:from-a} cannot be shown by this method.
Indeed, none of the four orderings\\ $\mathtt{from(X, scons(X,Y))}$
$<_l \mathtt{from(s(X),Y)}$,  $\mathtt{from(s(X),Y)}
<_l \mathtt{from(X, scons(X,Y))}$,\\  $\mathtt{from(X, scons(X,Y))}$
$<_p \mathtt{from(s(X),}$ $\mathtt{Y)}$, and $ \mathtt{from(s(X),Y)} $
$<_p \mathtt{from(X, scons(X,Y))}$\\ holds because the subterms pairwise
disagree on the ordering. This situation is common for LP, where some
arguments hold input data and some hold output data, so that some
decrease while others increase in recursive calls. Nevertheless, $P_6$
{\em is} universally observable, and we want to be able to infer
this. Studying the S-resolution derivation for $P_6$ in
Section~\ref{sec:intro}, we note that universal observability of $P_6$
is guaranteed by contraction of \verb|from|'s second argument. It is
therefore sufficient to establish that terms get smaller in only one
argument. 
This inspires our
definition of a \emph{contraction ordering}, which takes advantage of the tree representation of terms.

\begin{definition}[Contraction, recursive contraction]\label{df:rm}
If $t_1$ and $t_2$ are terms, then $t_2$ is a {\em
  contraction} of $t_1$ (written $t_1 \triangleright t_2$) if there is a leaf node $t_2(w)$ on
a branch $B$ in $t_2$, and there exists a branch $B'$ in $t_1$ that is identical to $B$ up to node $w$, 
however, $t_1(w)$ is not a leaf.
If, in addition,
$\mathit{subterm}(t_1,w)$ contains the symbol given by $t_2(w)$, then
$t_2$ is a \emph{recursive contraction} of $t_1$.

We distinguish \emph{variable contractions} and
\emph{constant contractions} according as $t_2(w)$ is
a variable or constant, and call $\mathit{subterm}(t_1,w)$ a
\emph{reducing subterm} for $t_1 \triangleright t_2$ at node
$w$. We call $\mathit{subterm}(t_1,w)$ a \emph{recursive, variable or constant
reducing subterm} if $t_1 \triangleright t_2$ is a recursive, variable or constant contraction, respectively.
\end{definition}


\begin{example}[Contraction orderings]
We have 
 $\mathtt{from(X,
scons(X,Y))} \triangleright \mathtt{from(s(X),Y)}$, as the leaf $\mathtt{Y}$ in the latter is ``replaced" by the term $\mathtt{scons(X,Y)}$ in the former.
Formally, $\mathtt{scons(X,Y)}$ is a recursive and variable reducing subterm.
It can be used to
certify termination of all rewriting derivations for $P_6$. Note that
$\mathtt{from(s(X),Y)} \triangleright \mathtt{from(X, scons(X,Y))}$
also holds, with (recursive and variable) reducing subterm $\mathtt{s(X)}$.
\end{example}

\noindent The fact that $\triangleright$ is not well-founded makes
 reasoning
about termination 
delicate.  Nevertheless, contractions emerge as precisely the
additional ingredient needed to formulate our productivity check
for a sufficiently general and interesting class of logic programs.



In general, static termination checking for LP suffers serious
limitations; see, e.g.,~\cite{deSchreye1994199}. The
following example illustrates this phenomenon.

\begin{example}[Contraction ordering on clause terms is insufficient for termination checks]\label{ex:not-simple}
The program $P_{7}$, that is not universally observable, is given by mutual recursion:

\noindent
0. $\mathtt{p(s(X1),X2,Y1,Y2)} \, \gets \,
 \mathtt{q(X2,X2,Y1,Y2)}$\\ 
1. $\mathtt{q(X1,X2,s(Y1),Y2)} \, \gets \,
\mathtt{p(X1,X2,Y2,Y2)}$

\noindent
No two terms from the same clause of $P_7$ can be related by any
contraction ordering because their head symbols differ. But recursion
arises for $P_7$ when a derivation calls its two clauses alternately,
so we would like to examine rewriting derivations for queries, such as
$? \gets \mathtt{p(s(X1),X2,s(Y1),Y2)} $ and
$? \gets \mathtt{p(s(X1),s(X2),s(Y1),s(Y2))}$, that exhibit its
recursive nature. Unfortunately, such queries are not given directly
by $P_7$'s syntax, and so are not available for static program
analysis.
\end{example}

As  static checking for contraction ordering in clauses is not sufficient, we will define dynamic checks in the next section.
The idea is to build a rewriting tree for each clause, and check whether term trees featured  in that derivation tree obey contraction ordering.

\vspace*{-0.15in}

\section{Rewriting Trees: Guardedness Checks for Rewriting
Derivations}\label{sec:cotrees}  

To properly reason about rewriting derivations in LP, we need to take
into account that i) in LP, unlike, e.g., in TRS, we have conjuncts of
terms in the bodies of clauses, and ii) a logic program can have
overlapping clauses, i.e., clauses whose heads unify.  These two facts
have been analysed in detail in the LP literature, usually using the
notion of and-or-trees and, where optimisation has been concerned,
and-or-parallel trees. We carry on this tradition and consider a
variant of and-or trees for derivations. However, the trees we
consider are not formed by general SLD-resolution, but rather by term
matching resolution. {\em Rewriting trees} are so named because each
of their edges represents a term matching resolution step, i.e., a
matching step as in term rewriting.

\begin{definition}[Rewriting tree]\label{def:cointree}
Let $P$ be a logic program with $n$ clauses, and $A$ be an atomic formula. The \emph{rewriting
tree} for $P$ and $A$ is the possibly infinite tree $T$ satisfying the following properties.
\begin{itemize}
\item $A$ is the root of $T$
\item Each node in $T$ is either an and-node or an or-node
\item Each or-node is given by $P(i)$, for some $i \in \{0, \ldots ,n\}$
\item Each and-node is an atom seen as a term tree.
\item For every and-node $A'$ occurring in $T$, if there exist exactly $k>0$ distinct clauses
$P(j), \ldots ,P(m)$ in $P$ (a clause $P(i)$ has the form $B_i \gets  B^i_1,\ldots ,B^i_{n_i}$ for some $n_i$), such that
$A' = \theta_j(B_j) = \ldots  = \theta_m(B_m)$, for mgms $\theta_j,\ldots ,\theta_m$, then $A'$ has exactly $k$ children given
by or-nodes $P(j), \ldots ,P(m)$, such that, every or-node $P(i)$ has $n_i$ children given by and-nodes
$\theta_i(B^i_1), \ldots , \theta_i(B^i_{n_i})$.
\end{itemize}
\end{definition}

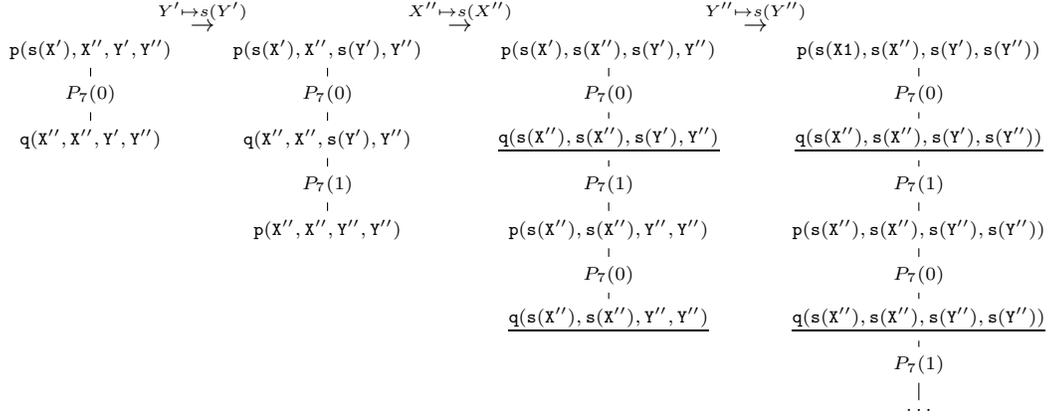
\begin{figure*}[t]
\begin{center}
\vspace*{-0.1in}\hspace*{-0.45in}
\begin{tikzpicture}[scale=.8,font=\scriptsize, baseline=(current bounding box.north),grow=down, level distance=7.5mm, level 1/.style={sibling distance=45mm},
level 2/.style={sibling distance=45mm},
level 3/.style={sibling distance=15mm}]
\node (root){$\mathtt{p(s(X'),X'',Y',Y'')}$}
child { node {$P_7(0)$}
child { node {$\mathtt{q(X'',X'',Y',Y'')}$}
}};
\end{tikzpicture}
\hspace*{-0.2in}
$\stackrel{Y' \mapsto s(Y')}{\rightarrow}$
\hspace*{-0.2in}
\begin{tikzpicture}[scale=.8,font=\scriptsize, baseline=(current bounding box.north),grow=down, level distance=7.5mm, level 1/.style={sibling distance=45mm},
level 2/.style={sibling distance=45mm},
level 3/.style={sibling distance=45mm}]
\node (root) {$\mathtt{p(s(X'),X'',s(Y'),Y'')}$}
child { node{$P_7(0)$}
child { node {$\mathtt{q(X'',X'',s(Y'),Y'')}$}
child { node{$P_7(1)$}
child { node {$\mathtt{p(X'',X'',Y'',Y'')}$}
}}
}};
\end{tikzpicture}
\hspace*{-0.2in}
$\stackrel{X'' \mapsto s(X'')}{\rightarrow}$
\hspace*{-0.2in}
\begin{tikzpicture}[scale=.8,font=\scriptsize, ,baseline=(current bounding box.north),grow=down,level distance=7.5mm, level 1/.style={sibling distance=45mm},
level 2/.style={sibling distance=45mm},
level 4/.style={sibling distance=45mm},
level 5/.style={sibling distance=45mm},
level 3/.style={sibling distance=45mm}]
\node (root) {$\mathtt{p(s(X'),s(X''),s(Y'),Y'')}$}
child { node{$P_7(0)$}
child { node {\underline{$\mathtt{q(s(X''),s(X''),s(Y'),Y'')}$}}
child { node{$P_7(1)$}
child { node {$\mathtt{p(s(X''),s(X''),Y'',Y'')}$}
child { node{$P_7(0)$}
child { node
{\underline{$\mathtt{q(s(X''),s(X''),Y'',Y'')}$}}
}}
}}}};
\end{tikzpicture}
\hspace*{-0.2in}
$\stackrel{Y'' \mapsto s(Y'')}{\rightarrow}$
\hspace*{-0.2in}
\begin{tikzpicture}[scale=.8,font=\scriptsize, level distance=7.5mm,  ,baseline=(current bounding box.north),grow=down, level 1/.style={sibling distance=45mm},
level 2/.style={sibling distance=45mm},
level 4/.style={sibling distance=45mm},
level 5/.style={sibling distance=45mm},
level 3/.style={sibling distance=45mm}]
\node (root) {$\mathtt{p(s(X1),s(X''),s(Y'),s(Y''))}$}
child { node{$P_7(0)$}
child { node {\underline{$\mathtt{q(s(X''),s(X''),s(Y'),s(Y''))}$}}
child { node{$P_7(1)$}
child { node {$\mathtt{p(s(X''),s(X''),s(Y''),s(Y''))}$}
child { node{$P_7(0)$}
child { node
{\underline{$\mathtt{q(s(X''),s(X''),s(Y''),s(Y''))}$}}
child { node{$P_7(1)$}
child { node{$\ldots$}}}
}}
}}
}};
\end{tikzpicture}
 \vspace*{-0.08in}
\caption{\footnotesize{An initial fragment of the derivation tree (comprising four rewriting trees)
    for the program $P_{7}$ of
    Example~\ref{ex:not-simple} and the atom $\mathtt{p(s(X'),X'',Y',Y'')}$. Its third and fourth rewriting trees
    each contain an unguarded loop (underlined), so both are
    unguarded. The fourth tree is infinite.}\vspace*{-0.2in}
    }\label{pic:infder2}
\end{center}
\end{figure*}

\noindent 
When constructing rewriting trees, we assume a suitable algorithm~\cite{JKK15} for
renaming free variables in clause bodies apart. Figure~\ref{pic:infder2} gives examples of rewriting trees.
An and-subtree of a rewriting tree (a subtree in which a derivation always pursues only one or-choice at a time)   is a \emph{rewriting derivation}, see~\cite{JKK15}
for a formal definition.

\vspace*{-0.01in}

Because mgms are unique up to variable renaming, given  a program $P$ and an atom $A$,
rewriting tree $T$ for $P$ and $A$ {\em is} unique.  
Following the same principle as with definition of term trees, we use suitably defined finitely-branching tree languages for indexing  rewriting trees, see~\cite{JKK15} for precise definitions.
When we need to talk about a node of a rewriting tree $T$ indexed by a word $w \in L$, we will use notation $T(w)$.

We can now formally define our notion of universal observability.
\begin{definition}[Universal observability]\label{df:prod}
A program $P$ is \emph{universally observable} if, for
every atom $A$, 
the rewriting tree for $A$ and $P$ is finite.
\end{definition}
Programs $P_1$, $P_1'$, $P_2$, $P_5$, $P_6$ are universally
observable, whereas programs $P_3$, $P_4$ and $P_7$ are
not. 
An exact analysis of why $P_7$ is not universally
observable is given in Example~\ref{ex:not-simple2}.

We can now apply the contraction ordering we defined in the previous
section to analyse termination properties of rewriting trees.  A
suitable notion of guardedness can be defined by checking for
loops in rewriting trees whose terms fail to decrease by any
contraction ordering.  But note that our notion of a loop is more
general than that used in CoLP~\cite{GuptaBMSM07,SimonBMG07} since
it does not require the looping terms to be unifiable.

\begin{definition}[Loop in a rewriting tree]\label{def:loop}
Given a program $P$ and an atom $A$ 
the rewriting tree $T$ for $P$ and $A$
 \emph{contains a loop} at nodes $w$ and $v$,
denoted $\mathit{loop}(T,w,v)$, if $w$ properly precedes $v$ on some
branch of $T$, $T(w)$ and $T(v)$ are and-nodes whose atoms have the
same predicate, and parent or-nodes of $T(w)$ and $T(v)$ are given by the same clause $P(i)$. 
\end{definition}

\noindent Examples of loops in rewriting trees are given (underlined) in Figure~\ref{pic:infder2}.

If $T$ has a loop at nodes $w$ and $v$, and if $t$ is a recursive
reducing subterm for $T(w) \triangleright
T(v)$, then $\mathit{loop}(T,w,v)$ is \emph{guarded}
by $(P(i), t)$, where $P(i)$ is the clause that was resolved against to obtain $T(w)$ and $T(v)$. It is \emph{unguarded} otherwise. A rewriting tree $T$
is {\em guarded} if all of its loops are guarded, and is {\em
unguarded} otherwise. We write $GC2(T)$ when $T$ is guarded, and say
that 
$GC2(T)$
holds. 

\begin{example}
In Figure ~\ref{pic:infder2}, we have (underlined) loops in the third rewriting tree (for $\mathtt{q(s(X''),s(X''),s(Y'),Y'')}$ and $\mathtt{q(s(X'),s(X''),Y'',Y'')}$) and the fourth rewriting tree
(for $\mathtt{q(s(X''),s(X''),s(Y'),s(Y''))}$ and $\mathtt{q(s(X''),s(X''),s(Y''),s(Y''))}$). Neither is guarded.  In the former case, there is a contraction on the third argument,
 but because $\mathtt{s(Y')}$ and $\mathtt{Y''}$
do not share a variable, it is not recursive contraction. In the latter loop, there is no contraction at all.
\end{example}

\noindent By
Definition~\ref{def:loop}, each repetition of a clause and predicate in a branch of a rewriting tree triggers a check to see
if the loop is guarded by some recursive reducing subterm.
\begin{proposition}[$GC2$ is decidable]\label{prop:ct-guard-dec}
GC2 is a decidable property of rewriting trees.\footnote{All proofs
are in an \textbf{Appendix A} supplied as supplementary material online. Corresponding pseudocode algorithms are given in \textbf{Appendix B}. }
\end{proposition}
\noindent
The proof of Proposition~\ref{prop:ct-guard-dec} also establishes that
every guarded rewriting tree is finite. 

The decidable guardedness property $GC2$ is a property of individual rewriting
trees. But our goal is to decide guardedness universally, i.e., for
{\em all} of a program's rewriting trees.  The next example shows that
extrapolating from existential to universal guardedness is
a difficult task.

\begin{example}[Existential guardedness does not imply universal guardedness]\label{ex:not-simple2} 
For program $P_{7}$, the rewriting trees constructed for the two clause heads $\mathtt{p(s(X'),X'',Y',Y'')}$ and\\ $\mathtt{q(s(X'),X'',s(Y'),Y'')}$  are both guarded since
neither contains any loops at all. Nevertheless, there is a rewriting
tree for $P_7$ (the last tree in Figure~\ref{pic:infder2}) that is
unguarded and infinite. 
The third tree is not guarded (due to the unguarded loop), but it is finite.
\end{example}

\noindent The example above shows that our initial idea of checking rewriting trees generated by clause heads is insufficient to detect all cases of 
nonterminating rewriting.
Since a similar situation can obtain for any finite set of rewriting
trees, universal observability, and hence observational productivity,
of programs cannot be determined by guardedness of rewriting trees for
program clauses alone. The next section addresses this problem.

\vspace*{-0.1in}

\section{Derivation Trees: Observational Productivity
Checks}\label{sec:guardedness} 

\vspace*{-0.1in} 

The key idea of this section is, given a program $P$, to identify a finite set $S$ of rewriting trees for $P$ such that checking guardedness of 
all rewriting trees in $S$
is sufficient for guaranteeing guardedness of \emph{all} rewriting trees for $P$.
One way to identify such sets will be to use the strategy of Example~\ref{ex:not-simple2} and Figure~\ref{pic:infder2}:
for every  clause $P(i)$ of $P$, to construct a rewriting tree for the head of $P(i)$, and, if that tree is guarded,
 explore what kind of mgus the leaves of that tree generate, and see if applications of those mgus may give an unguarded tree.
As Figure~\ref{pic:infder2} shows, we may need to apply this method iteratively until we find a nonguarded rewriting tree. But we want the number of such iterations to be finite.
This section presents a solution to this problem.
 
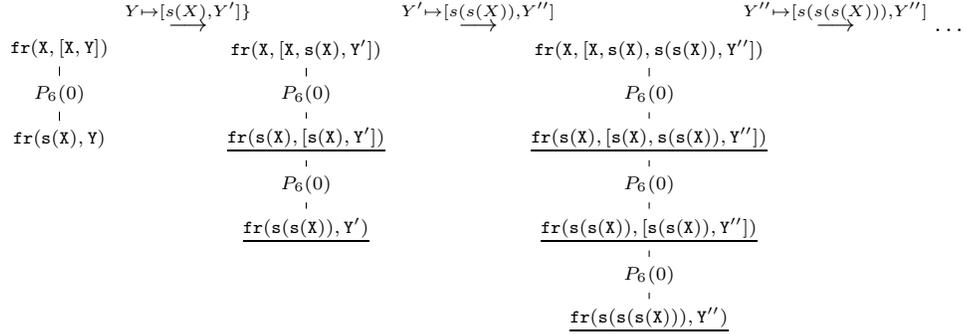
\begin{figure*}[t]
\vspace*{-0.1in}\hspace*{-0.1in}
\begin{tikzpicture}[scale=0.30,baseline=(current bounding box.north),grow=down,level distance=20mm,sibling distance=50mm,font=\scriptsize]
  \node { $\mathtt{fr(X,[X,Y])}$}
	child { node{ $P_6(0)$}
	child { node{ $\mathtt{fr(s(X),Y)}$ }}};
  \end{tikzpicture}~$\stackrel{Y \mapsto [s(X),Y'] \}}{\longrightarrow}$\hspace*{-0.21in}
	\begin{tikzpicture}[scale=0.30,baseline=(current bounding box.north),grow=down,level distance=20mm,sibling distance=60mm,font=\scriptsize ]
  \node { $\mathtt{fr(X,[X, s(X),Y'])}$}
		child { node{ $P_6(0)$}
	child { node{ $\mathtt{\underline{fr(s(X),[s(X),Y'])}}$ }
		child { node{ $P_6(0)$}
            child { node {$\mathtt{\underline{fr(s(s(X)),Y')}}$}}}}};
  \end{tikzpicture}~$\stackrel{Y' \mapsto [s(s(X)),Y'']}{\longrightarrow}$\hspace*{-0.21in}
	\begin{tikzpicture}[scale=0.30,baseline=(current bounding box.north),grow=down,level distance=20mm,sibling distance=60mm,font=\scriptsize]
  \node { $\mathtt{fr(X,[X, s(X), s(s(X)),Y''])}$}
	child { node{ $P_6(0)$}
	child { node{ $\mathtt{\underline{fr(s(X),[s(X),s(s(X)),Y''])}}$ }
	 child { node{ $P_6(0)$}
          child { node {$ \mathtt{\underline{fr(s(s(X)),[s(s(X)),Y''])}}$}
child { node{ $P_6(0)$}
          child { node {$\mathtt{\underline{fr(s(s(s(X))),Y'')}}$}}}}}}};
  \end{tikzpicture}\hspace*{-0.21in}
$\stackrel{Y'' \mapsto [s(s(s(X))),Y'']}{\longrightarrow} \ldots$
\caption{\footnotesize{An initial fragment of the infinite derivation
    tree $D$ for the program $P_6$  from
    Example~\ref{ex:from-a} and its clause head. It is also the observation subtree of $D$. We abbreviate $\mathtt{scons}$ by $[,]$,
    and $\mathtt{from}$ by $\mathtt{fr}$. The guarded loops in each of
    its rewriting trees are underlined.}\vspace*{-0.1in}
    }\label{pic:GC}
\end{figure*}

We start with a formal definition of rewriting tree transitions, which we have seen already in Figure~\ref{pic:infder2}, see also Figure~\ref{pic:GC}.

\vspace*{-0.05in}

\begin{definition}[Rewriting tree transition]\label{def:resapp}
 Let $P$ be a program and $T$ be a rewriting tree for $P$ and an atom $A$.
If $T(w)$ is a leaf node of $T$ given by an atom $B$, and $B$ unifies with a clause $P(i)$ via mgu $\sigma$,
we define a tree $T_{w}$ as follows: we apply $\sigma$ to every and-node of $T$, and extend the branches where required, 
according to Definition~\ref{def:cointree}.

 Computation of $T_w$ from $T$ is denoted 
 $T \rightarrow T_w$.  The operation $T \rightarrow T_w$ is the
\emph{tree transition} for $T$ and $w$.
\end{definition}

\vspace*{-0.05in}

\noindent If a rewriting tree $T$ is constructed for a program $P$ and an atom $A$,
 a (finite or infinite)
sequence $T 
\rightarrow T' \rightarrow T'' \rightarrow \ldots$ of tree
transitions  is an {\em $S$-resolution derivation} for
 $P$ and $A$.  
For a given rewriting tree $T$, several different S-resolution derivations are possible from $T$. This gives rise to the notion of a derivation tree.

\vspace*{-0.05in}

\begin{definition}[Derivation tree, guarded derivation tree]\label{df:CD}
Given a logic program $P$ and an atom $A$, the \emph{derivation tree}
$D$ for $P$ and $A$ is defined as follows:
\begin{itemize}
	\item The root of $D$ is given by the rewriting tree for $P$ and $A$.
	\item For a rewriting tree $T$ occurring as a node of $D$,
	if there exists a transition $T 
\rightarrow T_w$, for some leaf node $w$ in $T$, then the node $T$ has a child given by $T_w$.
\end{itemize}
A derivation tree is {\em guarded} if each of its nodes is a guarded
rewriting tree, i.e., if $GC2(T)$ holds for each of its nodes $T$.
\end{definition}

\vspace*{-0.05in}

\noindent Figure~\ref{pic:infder2} shows an initial fragment of the derivation tree for $P_7$ and $\mathtt{p(s(X'),X'',Y',Y'')}$.

Note that we now have three kinds of trees: term trees have signature symbols as nodes, rewriting trees have atoms (term trees) as nodes, and derivation trees have rewriting trees as nodes.
For a given $P$ and $A$, 
the derivation tree for $P$ and $A$ is unique up to renaming. 
We use our usual notation $D(w)$ to refer to the node of $D$ at index $w \in L$. 


\vspace*{-0.05in}

\begin{definition}[Existential liveness, observational
productivity]\label{df:prod2} Let $P$ be
a universally observable program  and let $A$ be an atom. 
 An
S-resolution derivation for $P$ and $A$ is \emph{live} if it
 constitutes an infinite branch
of the derivation tree for $P$ and $A$.  The program $P$ is \emph{existentially live} if there exists
a live S-resolution derivation for $P$ and some atom $A$. $P$ is \emph{observationally
productive} if it is universally observable and existentially
live.
\end{definition}

\vspace*{-0.05in}

To show that observational productivity is
semi-decidable, we first show that universal observability is
semi-decidable by means of a finite (i.e., decidable) guardedness
check. We started this section by motivating the need to construct a finite  set $S$ of rewriting trees checking guardedness of which will
guarantee guardedness for \emph{any} rewriting tree for the given program.
Our first logical step is to use derivation trees built for clause heads as generators of such a set $S$.
Due to the properties of mgu's used in forming branches of derivation trees,  derivation trees
constructed for  clause heads 
generate the set of \emph{most general} rewriting trees. The next lemma exposes this fact: 

\begin{lemma}[Guardedness of  derivation trees implies universal observability]\label{lem:main1} 
Given a program $P$, if derivation trees for $P$ and each $\mathit{head}(P(i))$ are guarded, then $P$ is universally
observable.
\end{lemma}

\vspace*{-0.05in}


However, derivation trees are infinite, in general. So it still remains to define a method that extracts representative finite subtrees from such derivation trees;
we call such subtrees \emph{observation subtrees}. For this, we need only be able to detect an invariant property
guaranteeing guardedness through tree transitions in the given derivation tree. 
 To
illustrate, let us check guardedness of the program $P_6$. 
As it consists of just one clause, we take the head of that clause as the goal atom, and start constructing
 the infinite derivation tree $D$ for $P_6$ and $\mathtt{from(X, scons(X,Y))}$ as shown in Figure~\ref{pic:GC}.
The first rewriting tree in the derivation tree has no loops, so we cannot identify any invariants.
We make a transition to the second rewriting tree which has one loop (underlined) involving the recursive reducing subterm $\mathtt{[s(X),Y']}$.
This reducing subterm is our first candidate invariant, it is the pattern that is \emph{consumed} from the root of the second rewriting tree to its leaf. 
We now need to check this pattern is added back, or \emph{produced}, in the next tree transition. The next 
mgu involves substitution $\mathtt{Y' \mapsto [s(s(X)),Y'']}$. Because this derivation gradually computes an infinite irrational term (rational terms are terms that can
be represented as trees that have a finite number of distinct
subtrees),  the two 
terms $\mathtt{[s(X),Y']}$ and $\mathtt{[s(s(X)),Y'']}$ we have identified are not unifiable. We need to be able to abstract away from their current shape and identify a common pattern, which  is $\mathtt{[\_,\_]}$. 
By the properties of mgu's used in transitions, such most general pattern can always be extracted from the clause head itself. 
Indeed, the subterm of the clause head $\mathtt{from(X, scons(X,Y))}$ has the subterm $\mathtt{[X,Y]}$ that is exactly the pattern we look for.
Thus, our current \emph{(coinductive) assumption} is: \emph{given a rewriting tree $T$ in the derivation tree $D$, 
 $\mathtt{[X,Y]}$ will be {\em consumed} by rewriting
steps from its root to its leaves, and exactly
$\mathtt{[X,Y]}$ will be {\em produced} (i.e., added back) in the next
tree transition.} $\mathtt{X}$ and $\mathtt{Y}$ are seen as placeholders for some terms. Consumption is always finite (by the loop guardedness), and production is potentially infinite. 

We now need to check that this coinductive assumption will hold for the next rewriting tree of $D$. 
The third rewriting tree indeed 
has guarded loops with recursive reducing subterm $\mathtt{[s(s(X)),Y'']}$, and the next mgu it gives rise to is $\mathtt{Y'' \mapsto [s(s(s(X))),Y'']}$.
Again, to abstract away the common pattern, we look for a subterm in the clause head of $P_6(0)$ that matches with both of these terms,
 it is the same subterm $\mathtt{[X,Y]}$. Thus, our coinductive assumption holds again, and we conclude by coinduction that the same pattern will hold for any further rewriting tree in $D$.
When implementing this reasoning, we take the \emph{observation subtree} of $D$ up to the third tree shown in Figure~\ref{pic:GC} as a 
sufficient set of rewriting trees to check guardedness of (otherwise infinite) $D$.

The rest of this section generalises and formalises this approach.
In the next definition, we introduce the notion of a \emph{clause projection} to talk about the process of ``abstracting away" a pattern
from an mgu $\sigma$ by matching it with a subterm $t$ of a clause head.
When $t$ also matches with a recursive reducing subterm of a loop in a rewriting tree, we call $t$ a \emph{coinductive invariant}.


\begin{definition}[Clause projection and coinductive invariant]\label{def:gc} Let $P$ be a program and
$A$ be an atom, and let $D$ be a
derivation tree for $P$ and $A$ in which a tree transition from $T$
to $T'$ is induced by an mgu $\sigma$ of some
$P(k)$ and an atom $B$ given by a leaf node $T(u)$. 

The \emph{clause projection} for $T'$, denoted $ \pi(T')$, is the
set of all triples $(P(k),t,v)$, where $t$ is a subterm of $\mathit{head}(P(k))$ at position $v$,
such that the following conditions hold:
	$\sigma(B) \triangleright B$
with variable reducing subterm $t'$, and $t'$ matches against $t$ (i.e.  $t' = \sigma'(t)$ for some $\sigma'$).  

Additionally,
the \emph{coinductive invariant} at $T'$, denoted $\kw{ci}(T')$,
 is a subset of  the \emph{clause projection} for $T'$,
satisfying the following condition.
An element $(P(k),t,v) \in \pi(T')$ is also in $\kw{ci}(T')$, if $T$ contains a loop in the branch
leading from $T$'s root to $T(u)$ that is guarded by $(P(k),t'')$ for
some $t''$ such that  $t''$ matches against $t$ ($t'' = \theta(t)$ for some $\theta$). 

Given a program $P$, an atom $A$ and a derivation tree $D$ for $P$ and $A$,
the \emph{clause projection set} for $D$ is
$\kw{cproj}(D) \,=\, \bigcup_T \pi(T)$ and the {\em coinductive invariant
set} for $D$ is $\kw{cinv}(D) \,=\,\bigcup_T \kw{ci}(T)$, where these
unions are taken over all rewriting trees $T$ in $D$.
\end{definition}

\begin{example}[Clause projections and coinductive invariants]\label{ex:from2}
Coming back to 
Figure~\ref{pic:GC}, 
the mgu for the first transition
is $\sigma_1 = \{\mathtt{X'} \mapsto
\mathtt{s(X)}, \mathtt{Y}\mapsto \mathtt{scons(s(X),Y')}\}$ 
(renaming of variables in $P_6(0)$ with primes), that for the second
is $\sigma_{2} = \{\mathtt{X''} \mapsto
\mathtt{s(s(X))}, \mathtt{Y'}\mapsto \mathtt{scons(s(s(X)),Y'')}\}$
(renaming of variables in $P_6(0)$ with double primes), etc.  
Clause projections are given by $\pi(T) = \{(P_6(0),
\mathtt{scons(X,Y)},1)\}$ for all trees $T$ in this derivation, and thus $\kw{cproj}(D)$ is the finite set.  Moreover, for the first rewriting tree $T$,
$\kw{ci}(T) = \emptyset$, and $\kw{ci}(T') = \{(P_6(0),$
$\mathtt{scons(X,Y)},1)\}$ for all trees $T'$ except for the first one, so $\kw{cinv}(D) =
\{(P_6(0), \mathtt{scons(X,Y)},1)\}$ is the finite set too.
\end{example}

\noindent The clause projections for the derivation of
Figure~\ref{pic:infder2} are $\pi(T') = \pi(T''') = (P(1),$
$\mathtt{s(Y1)}, 2)$, and $\pi(T'') = (P(0), \mathtt{s(X1)},
0)$, where $T', T'', T'''$ refer to the second, third and fourth rewriting tree of that derivation.
All coinductive invariants for that derivation are empty, since none of
these rewriting trees contain guarded loops. 

Generally, clause projection sets are finite, as the number of subterms in the clause heads of $P$ is finite. This property is crucial for
termination of our method:

\vspace*{-0.05in}
\begin{proposition}[Finiteness of clause projection sets]\label{prop:finite}
Given a program $P$, an atom $A$, and a derivation tree $D$ for $P$ and $A$,
the clause projection set $\kw{cproj}(D)$ is finite.
\end{proposition}
\vspace*{-0.1in}
In particular, this holds for derivation trees induced by  clause heads.

We terminate the construction of each branch of a derivation tree when we notice repeating coinductive invariant.
A subtree we get as a result is an observation subtree. Formally,
given a derivation tree $D$ for a program $P$ and an atom $A$, with a branch in which nodes $D(w)$ and $D(wv)$ are defined,
if $\kw{ci}(D(w))
= \kw{ci}(D(wv)) \neq \emptyset$, then $D$ has a {\em guarded
transition} from $D(w)$ to $D(wv)$ (denoted $D(w) \Longrightarrow
D(wv)$). Every guarded transition thus identifies a repeated
``consumer-producer" invariant in the derivation from $D(w)$ to $D(wv)$. This tells us 
that observation of this branch of $D$ can
 be concluded. Imposing this condition on all branches of $D$
 gives us  a general method to construct finite observation subtrees of potentially infinite derivation trees: 

\vspace*{-0.05in}

\begin{definition}[Observation subtree of a derivation tree]\label{def:resapp3}
If $D$ is  a derivation tree for a program $P$ and an atom $A$, the tree $D'$ is the {\em observation subtree} of
$D$ if\\ 
1) the roots
of $D$ and $D'$ are given by the rewriting tree for $P$ and $A$, and \\
2) if $w$ is a
node in both $D$ and $D'$, then the rewriting trees in $D$ and $D'$ at
node $w$ are the same and, for every child $w'$ of $w$ in $D$, the
rewriting tree of $D'$ at node $w'$ exists and is the same as the
rewriting tree of $D$ at $w'$, unless either\\ a) GC2 does not hold for $D(w')$, or\\
b) there exists a $v$ such that $D(v) \Longrightarrow D(w)$.\\  In
either case, $D'(w)$ is a leaf node. 
We say that $D'$ is {\em
unguarded} if Condition 2a holds for at least one of $D$'s nodes, and
that $D'$ is {\em guarded} otherwise.
\end{definition}

A branch in an observation subtree is thus truncated when it reaches
an unguarded rewriting tree or its coinductive invariant repeats. The observation subtree of any derivation tree is unique.
The following proposition and lemma prove the two most crucial properties of observation subtrees: that they are always finite, and that checking their guardedness is sufficient for establishing guardedness of the whole derivation trees.

\begin{proposition}[Finiteness of observation subtrees]\label{prop:fot}
If $D$ is  a derivation tree for a program $P$ and an atom $A$
then the observation subtree of $D$ is finite.
\end{proposition}
\vspace*{-0.1in}

\begin{lemma}[Guardedness of observation subtree implies guardedness
of derivation tree]
\label{lem:main2}
If the observation subtree for  a derivation tree $D$ is guarded,
then $D$ is guarded.
\end{lemma}

\begin{example}[Finite observation subtree of an infinite derivation tree]\label{ex:from-guarded}
The initial fragment $D'$ of the infinite derivation tree $D$ given by the three rewriting trees in
Figure~\ref{pic:GC} is $D$'s observation subtree. The third rewriting tree $T''$ in $D$ is the last node
in the observation tree $D'$ because $\kw{ci}(T')
= \kw{ci}(T'') = \{(P_6(0),$ $\mathtt{scons(X,Y)},$ $1)\}
\not = \emptyset$. Since $D'$ is guarded, Lemma~\ref{lem:main2} above
ensures that the whole infinite derivation tree $D$ is guarded.
\end{example}


It now only remains to put the properties of the observation subtrees into practical use, and, given a program $P$, construct finite observation subtrees for each of its clauses.
If none of these observation subtrees detects unguarded rewriting trees, we have guarantees that this program will never give rise to infinite rewriting trees. The next definition, lemmas and a theorem make this intuition precise. 

\begin{definition}[Guarded clause, guarded program]\label{df:G3}
Given a program $P$, its clause $P(i)$ is
{\em guarded} if the observation subtree for the derivation tree for $P$ and the atom $\mathit{head}(P(i))$ is
guarded, and $P(i)$ is {\em unguarded} otherwise. A program $P$ is {\em guarded}
if each of its clauses $P(i)$ is guarded, and {\em unguarded}
otherwise. We write $GC3(P(i))$ to indicate that $P(i)$ is guarded, and
similarly for $P$.
\end{definition}
Lemma~\ref{lem:GC3-t} uses Proposition~\ref{prop:fot} to show that GC3 is decidable. 
\begin{lemma}[GC3 is decidable]\label{lem:GC3-t}
$GC3$ is a decidable property of logic programs.
\end{lemma}


\vspace*{-0.1in}



\begin{theorem}[Universal observability is semi-decidable]\label{th:prod}
If $GC3(P)$ holds, then $P$ is universally observable.
\end{theorem}
\noindent
{\sc Proof:} If $GC3(P)$ holds, then the observation subtree for each
$P(i)$ is guarded. Thus, by Lemma~\ref{lem:main2}, the derivation tree for each
$P(i)$ is guarded. But then,  by Lemma~\ref{lem:main1}, $P$ is
universally observable.
Combining this with  Lemma~\ref{lem:GC3-t}, we also obtain that universal observability is semi-decidable.

\vspace*{0.05in}

\noindent The converse of Theorem~\ref{th:prod} does not hold: the program
comprising the clause $\mathtt{p(a)} \; \gets \; \mathtt{p(X)}$
is universally observable but not guarded, hence the above \emph{semi}-decidability result.

\vspace*{0.05in}

From our check for universal observability we obtain the desired
check for existential liveness, and thus for observational
productivity:
\begin{corollary}[Observational productivity is semi-decidable]
Let $P$ be a guarded logic program. If there exists a clause $P(i)$
such that the derivation tree $D$ for $P$ and $P(i)$ has an observation subtree $D'$ one
of whose branches was truncated by Condition 2b of
Definition~\ref{def:resapp3}, then $P$ is existentially live. In this case,
since $P$ is also guarded and hence universally observable, $P$ is observationally productive.
\end{corollary}

\vspace*{-0.1in}

\section{Related Work: Termination Checking in TRS and LP}\label{sec:trs}
\vspace*{-0.1in}

Because observational productivity is a combination of universal observability and existential liveness, and the former property amounts to termination of all rewriting trees,
there is an intersection between this work and termination checking in TRS~\cite{Terese,Arts2000,HM04}.

Termination checking via transformation of LP into TRS has been given in \cite{Schneider-KampGST06}.
Here we consider termination of restricted form of SLD-resolution (given by rewriting derivations),
therefore a much simpler method of translation of LP into TRS can be used for our purposes~\cite{FK16}:
Given a logic program $P$ and a  clause $P(i) = A \gets B_1, \ldots , B_n$ containing no existential variables, 
we define a rewrite rule $A \rightarrow f_i (B_1, \ldots, B_n)$ for some fresh function symbol $f_i$. 
Performing this translation for all clauses, we get a translation from $P$ to  a term-rewriting system $\mathcal{T}_P$.
  Rewriting derivations for $P$ 
	can be shown operationally equivalent to
term-rewriting reductions for $\mathcal{T}_P$; see~\cite{FK16} for a proof. 
Therefore, for logic programs containing no existential variables, any termination method from TRS may be applied to check universal observability (but not existential liveness).

Algorithmically, our guardedness check compares directly with the method of dependency pairs due to Arts and Giesl~\cite{Arts2000,HM04}.
Consider again $\mathcal{T}_P$ obtained from a program $P$.
The set $R$ of dependency pairs contains, for each rewrite rule $A \rightarrow f_i (B_1, \ldots, B_n)$ in $\mathcal{T}_P$, a pair  $(A,B_j)$, $j = 1, \ldots, n$; see~\cite{FK16}. 
The method of dependency pairs consists of checking whether there exists an infinite chain of dependency pairs $(s_i,t_i)_{i=1,2,3,\ldots}$ such that $\sigma_i(t_i) \rightarrow^* \sigma_{i+1}(s_{i+1})$.  
If there is no such infinite chain, then $\mathcal{T}_P$ is terminating. Again this translation from LP to dependency pairs in TRS is simpler than in~\cite{NguyenGSS07}, as rewriting derivations are a restricted form of SLD-resolution.
Due to the restricted syntax of  $\mathcal{T}_P$ (compared to the general TRS syntax), generating the set of dependency pairs is equivalent to generating a set of rewriting trees for each clause of $P$ and assuming $\sigma_i = \sigma_{i+1}$ (cf. our GC2).
To find infinite chains, a dependency graph is defined, in which dependency pairs are nodes and arcs are defined whenever a substitution that allows a transition from one pair to another can be found.
Finding such substitutions is the hardest part algorithmically. 
Note that every pair of neighboring and-nodes in a rewriting tree corresponds to a node in a dependency graph.
Generating arcs in a dependency graph is equivalent to using our GC3 to find a representative set of substitutions. 
However, the way GC3 generates such substitutions via rewriting tree transitions differs completely from the methods  approximating dependency graphs~\cite{Arts2000,Terese}, and relies on the properties of S-resolution, rather than recursive path orderings. 
This is because GC3 additionally generates coinductive invariants for checking existential liveness of programs.


Conceptually, observational productivity is a new property that does not amount to either termination or nontermination in LP or TRS. E.g. programs $P_3$ and $P_4$ are nonterminating (seen as LP or TRS), and  $P_8: p(X) \gets q(Y)$ is terminating (seen as LP and TRS) but none of them is productive. This is why the existing powerful tools (such as AProVE) and methods~\cite{Arts2000,HM04,NguyenGSS07,Schneider-KampGST06} that can check 
termination or nontermination in TRS or LP are not sufficient to serve as productivity checks.
To check \emph{termination} of rewriting trees, GC3 can be substituted by existing termination checkers for TRS, but 
none of the previous approaches can  semi-decide existential liveness as GC3 does. 


\vspace*{-0.1in}

\section{Implementation and Applications}\label{sec:app}
\vspace*{-0.1in}

We implemented the observational productivity checker in parallel Go
(golang.org)~\cite{Martin}, which allows to experiment with parallelisation of proof search~\cite{KSH14}. 
Loading a logic program $P$, one runs a command line to initialise the $GC3$ check.
The algorithm then 
certifies whether or not the program is guarded (and hence  universally observable).
If that is the case, it also checks whether $GC3$ found valid coinductive invariants, i.e. whether $P$ is existentially live and hence 
admits coinductive interpretations for some predicates.
	Appendix B (available in online  version) gives further details.

 

In the context of S-resolution~\cite{KPS12-2,JKK15}, observational productivity of a 
program is a pre-condition for (coinductive) soundness of S-resolution derivations. This gives the first application for the productivity checker.
But the notion of global productivity (as related to \emph{computations at infinity}~\cite{Llo88}) is  a general property tracing its roots to the 1980s. 
A program is productive, if it admits SLD- or S-resolution derivations that compute (or produce) an infinite term at infinity.
Thus the productivity checker has more general practical significance for Prolog.
In this paper we further exposed its generality  by  showing that productivity can be seen as a general property of logic programs, rather than property of
  derivations in some special dialect of Prolog.

Based on this observation, we identify three applications for productivity checks encompassing the S-resolution framework.
(1) In the context of CoLP~\cite{GuptaBMSM07,SimonBMG07} or any other similar tool based on loop detection in SLD-derivations,
one can run the observational productivity checker for a given program 
prior to running the usual interpreter of CoLP. If the program is certified as productive, all computations by CoLP for this program will be 
sound relative to the computations at infinity~\cite{Llo88}.
It gives a way to characterise a subset of theorems proven by CoLP that describe the process of \emph{production of infinite data}.
I.e., as explained in Introduction, CoLP will return answers for programs $P_3$, $P_4$ and $P_5$. But if we know that only $P_5$ is productive, we will know that only CoLP's answers for $P_5$ will correspond to production of  infinite terms at infinity.

\noindent (2) As our productivity checker also checks \emph{liveness} of programs, it effectively identifies which predicates may be given coinductive semantics.
This knowledge can be used to type predicates as inductive or coinductive. We can use these types to mark predicates in CoLP or any other 
coinductive dialect of logic programming, cf.~Appendix B.

\noindent (3) Observational productivity is also a guarantee that a sequence of mgus approximating the infinite answer 
 can be constructed \emph{lazily} even if the answer is irregular. E.g. our running example of program $P_6$ is irrational and hence cannot be handled by CoLP's loop detection. But even if we cannot form a closed-term answer for a query $\mathtt{from(0,X)}$, the productivity checker gives us a weaker 
but more general certificate that lazy approximation of our infinite answer is possible.

These three groups of  applications  show that the presented productivity checker can be implemented and applied in any dialect of logic programming, 
irrespective of  the fact that it initially arose from S-resolution research~\cite{KPS12-2,JKK15}. 

\vspace*{-0.1in}

\section{Conclusions}\label{sec:conclusion}

\vspace*{-0.1in}

In this paper we have introduced an observational counterpart to the
classical notion of global productivity of logic programs.  Using the
recently introduced formalism of S-resolution, we have defined
observational productivity as a combination of two program properties,
namely, universal observability and existential liveness.  We have
introduced an algorithm for semi-deciding observational productivity
for any logic program. We did not impose any restrictions on the syntax
of logic programs. In particular, our algorithm handles both existential
variables and non-linear recursion.

The algorithm relies on the observation that rewriting trees for
productive and guarded programs must show term reduction relative to a
contraction ordering from their roots to their leaves. But
S-resolution derivations involving such trees can only proceed by adding term
structure back in transitioning to new rewriting trees via mgus.  This
``producer/consumer'' interaction can be formally traced by observing
a derivation's coinductive invariants: these record exactly the term
patterns that both reduce in the loops of rewriting trees and are
added back in transitions between these trees. 


\vspace*{-0.1in}
\bibliographystyle{plain}
\bibliography{katya2}

\begin{thebibliography}{10}

\bibitem{Arts2000}
T.~Arts and J.~Giesl.
\newblock Termination of term rewriting using dependency pairs.
\newblock {\em TCS}, 236(1–2):133 -- 178, 2000.

\bibitem{Courcelle83}
B.~Courcelle.
\newblock Fundamental properties of infinite trees.
\newblock {\em TCS}, 25:95--169, 1983.

\bibitem{deSchreye1994199}
D~de~Schreye and S.~Decorte.
\newblock Termination of logic programs: the never-ending story.
\newblock {\em Journal of Logic Programming}, 19--20, Supplement 1:199--260,
  1994.

\bibitem{EndrullisGHIK10}
J.~Endrullis et~al.
\newblock Productivity of stream definitions.
\newblock {\em TCS}, 411(4-5):765--782, 2010.

\bibitem{EndrullisHHP015}
J.~Endrullis et~al.
\newblock A coinductive framework for infinitary rewriting and equational
  reasoning.
\newblock In {\em {RTA}}, pages 143--159, 2015.

\bibitem{FK16}
P.~Fu and E.~Komendantskaya.
\newblock Operational semantics of resolution and productivity in {H}orn clause
  logic.
\newblock {\em Formal Aspects of Computing}, 2016.

\bibitem{GuptaBMSM07}
G.~Gupta et~al.
\newblock Coinductive logic programming and its applications.
\newblock In {\em ICLP}, pages 27--44, 2007.

\bibitem{HM04}
N.N. Hirokawa and A.~Middeldorp.
\newblock Dependency pairs revisited.
\newblock In {\em RTA}, pages 249--268, 2004.

\bibitem{JKK15}
P.~Johann et~al.
\newblock Structural resolution for logic programming.
\newblock In {\em Technical Communications of ICLP}, 2015.

\bibitem{KSH14}
E.~Komendantskaya et~al.
\newblock Exploiting parallelism in coalgebraic logic programming.
\newblock {\em ENTCS}, (33):121--148, 2014.

\bibitem{KPS12-2}
E.~Komendantskaya et~al.
\newblock Coalgebraic logic programming: from semantics to implementation.
\newblock {\em Journal of Logic and Computation}, 26(2):745--783, 2016.

\bibitem{KJ15}
E.~Komendantskaya and P.~Johann.
\newblock Structural resolution: a framework for coinductive proof search and
  proof construction in {H}orn clause logic.
\newblock {\em Submitted}, 2015.

\bibitem{LeinoM14}
K.~R.M. Leino and M.~Moskal.
\newblock Co-induction simply - automatic co-inductive proofs in a program
  verifier.
\newblock In {\em {FM}}, pages 382--398, 2014.

\bibitem{Llo88}
J.W. Lloyd.
\newblock {\em Foundations of Logic Programming}.
\newblock Springer-Verlag, 2nd edition, 1988.

\bibitem{NguyenGSS07}
M.T. Nguyen et~al.
\newblock Termination analysis of logic programs based on dependency graphs.
\newblock In {\em LPOSTR 2007}, pages 8--22, 2007.

\bibitem{Pf92}
F.~Pfenning.
\newblock {\em Types in Logic Programming}.
\newblock The MIT Press, 1992.

\bibitem{ReynoldsB15}
A.~Reynolds and J.~Blanchette.
\newblock A decision procedure for (co)datatypes in {SMT} solvers.
\newblock In {\em CADE}, pages 197--213, 2015.

\bibitem{RohwedderP96}
E.~Rohwedder and F.~Pfenning.
\newblock Model and termination checking for higher-order logic programs.
\newblock In {\em ESOP}, pages 296--310, 1996.

\bibitem{Martin}
M.~Schmidt.
\newblock Productivity checker for {LP},
  \footnotesize{www.macs.hw.ac.uk/$\sim$ek19/CoALP/}, 2016.

\bibitem{Schneider-KampGST06}
P.~Schneider{-}Kamp et~al.
\newblock Automated termination analysis for logic programs by term rewriting.
\newblock In {\em LOPSTR}, pages 177--193, 2006.

\bibitem{SimonBMG07}
L.~Simon et~al.
\newblock Co-logic programming: Extending logic programming with coinduction.
\newblock In {\em ICALP}, pages 472--483, 2007.

\bibitem{Terese}
Terese.
\newblock {\em Term Rewriting Systems}.
\newblock Cambridge University Press, 2003.

\end{thebibliography}

\vfill

\pagebreak
\section*{Appendix A. Proofs}

\subsection*{Proof of Proposition~\ref{prop:ct-guard-dec}: $GC2$ is decidable}

Any rewriting tree $T$ is either finite or infinite.  If $T$ is
finite, then its guardedness is clearly decidable. So, we may, without loss of generality, assume $T$
is infinite.
 If $T$ is infinite
then it must have an infinite branch $B$. We now show that an
infinite branch $B$ in $T$ must necessarily contain an unguarded
loop. Thus, whether $T$ is finite or infinite, its guardedness is
decidable.

Assume $B$ has only guarded loops, but it is infinite. Since the
number of clauses and the number of function symbols in $\Sigma$ are
finite, $B$ must contain an infinite number of loops. Consider one
such infinite sequence $\mathtt{q}(t_{11},...,t_{1j}) \rightarrow
\ldots \rightarrow \mathtt{q}(t_{k1},...,t_{kj}) \rightarrow \ldots
\rightarrow \mathtt{q}(t_{l1},...,t_{lj}) \rightarrow \ldots$, where
$\mathtt{q}(t_{11},...,t_{1j})$, $\mathtt{q}(t_{k1},...,t_{kj})$,
$\mathtt{q}(t_{l1},...,t_{lj})$, ... are all atoms with the same predicate $\mathtt{q}$ obtained by rewriting using clause $P(i)$.
Because all loops in $B$ are guarded, we have
$\mathtt{q}(t_{11},...,t_{1j}) \triangleright
\mathtt{q}(t_{k1},...,t_{kj})$, $\mathtt{q}(t_{11},...,t_{1j})
\triangleright \mathtt{q}(t_{l1},...,t_{lj})$,... But since
$\mathtt{q}(t_{11},...,t_{1j})$ is finite, there are only finitely
many ways to construct a reducing subterm on it. Thus, there will
be a point when some terms $\mathtt{q}(t_{m1},...,t_{mj})$ and
$\mathtt{q}(t_{n1},...,t_{nj})$ in the infinite sequence in $B$ have
the same recursive reducing subterm $t^*$ relative to
$\mathtt{q}(t_{11},...,t_{1j})$.

Now, there are two cases: i) $\mathtt{q}(t_{11},...,t_{1j})
\triangleright \mathtt{q}(t_{m1},...,t_{mj})$ and
$\mathtt{q}(t_{11},...,t_{1j}) \triangleright
\mathtt{q}(t_{n1},...,t_{nj})$ hold, but
$\mathtt{q}(t_{m1},...,t_{mj}) \triangleright
\mathtt{q}(t_{n1},...,t_{nj})$ does not, and ii) the negation of this
case.
\begin{itemize}
\item If $\mathtt{q}(t_{11},...,t_{1j}) \triangleright
  \mathtt{q}(t_{m1}...t_{mj})$ and $\mathtt{q}(t_{11},...,t_{1j})
  \triangleright \mathtt{q}(t_{n1},...,t_{nj})$ hold, but not
  $\mathtt{q}(t_{m1},...,t_{mj}) \triangleright
  \mathtt{q}(t_{n1},...,t_{nj})$, then $\mathtt{q}(t_{m1},...,t_{mj})
  \triangleright \mathtt{q}(t_{n1},...,t_{nj})$ is an unguarded loop,
  which contradicts the assumption that all loops in $B$ are guarded.
\item If the negation holds --- i.e., if
  $\mathtt{q}(t_{m1},...,t_{mj}) \triangleright
  \mathtt{q}(t_{n1},...,t_{nj})$ holds or
  $\mathtt{q}(t_{11},...,t_{1j}) \triangleright
  \mathtt{q}(t_{m1},...,t_{mj})$ does not hold or
  $\mathtt{q}(t_{11},...,t_{1j}) \triangleright
  \mathtt{q}(t_{n1},...,t_{nj})$ does not hold --- then there are
  three cases. If either $\mathtt{q}(t_{11},...,t_{1j}) \triangleright
  \mathtt{q}(t_{m1},...,t_{mj})$ or $\mathtt{q}(t_{11},...,t_{1j})
  \triangleright \mathtt{q}(t_{n1},...,t_{nj})$ does not hold, then
  the existence of this unguarded loop in $B$ gives a
  contradiction. So we need only consider the case when
  $\mathtt{q}(t_{11},...,t_{1j}) \triangleright
  \mathtt{q}(t_{m1},...,t_{mj})$, $\mathtt{q}(t_{11},...,t_{1j})
  \triangleright \mathtt{q}(t_{n1},...,t_{nj})$, and
  $\mathtt{q}(t_{m1},...,t_{mj}) \triangleright
  \mathtt{q}(t_{n1},...,t_{nj})$ are all guarded loops in $B$. Let
  $t^{**}$ be the recursive reducing subterm for the loop
  $\mathtt{q}(t_{m1},...,t_{mj}) \triangleright
  \mathtt{q}(t_{n1},...,t_{nj})$.  Since the same recursive
  reducing subterm cannot be contracted twice along the same path
  from $\mathtt{q}(t_{11},...,t_{1j})$, we must have that $t^* \not =
  t^{**}$ and, moreover, contracting $t^{**}$ must somehow ``restore''
  $t^*$ to $\mathtt{q}(t_{11},...,t_{1j})$. And this means that $t^*$
  and $t^{**}$ must be ``independent", in the sense of being on
  independent paths in $\mathtt{q}(t_{11},...,t_{1j})$.  But then
  there will be cycles of terms in $B$ in which one argument of
  $\mathtt{q}$ decreases in one step and another independent one
  grows, and then the first argument grows while the other one
  decreases. So $\mathtt{q}(t_{11},...,t_{ij})$ will appear in the
  infinite branch infinitely many times (so definitely more than
  once!), and $B$ will thus contain an unguarded loop in this case as
  well. This is again a contradiction.
\end{itemize}

\subsection*{Proof of Lemma~\ref{lem:main1}: Guardedness of  derivation trees implies universal observability}

If $P$ is not universally observable, then there exists an atom $A$ such that 
the rewriting tree $T$ for $P$ and $A$ is
infinite. Moreover,
$A$ must match some clause $\mathit{head}(P(i))$ via a mgm $\theta$, so in fact $T$
is an infinite rewriting tree for $P$ and $\theta(\mathit{head}(P(i)))$,
 with additional condition that $\theta$ is also applied to all atoms of this tree.
 Then, as in the proof of Proposition~\ref{prop:ct-guard-dec}, there must exist
an unguarded loop $L$ on an infinite branch $B$ of $T$. 
We claim that, if we construct a derivation tree $D_i$ for the program $P$ and $\mathit{head}(P(i))$,
then some rewriting tree in $D_i$ will contain an
unguarded loop. Let us consider the construction of $D_i$.

If the first rewriting tree of $D_i$, i.e. the tree $T'$ for $P$ and $\mathit{head}(P(i))$
does not itself
contain an unguarded loop, then the branch in $T'$ corresponding to
$B$  in $T$ must have a leaf node $T'(w)$ given by an atom that unifies with a clause $P(k_1)$ via mgu $\sigma_1$, say. Moreover,
$P(k_1)$ is exactly the clause used to construct a node $T(wi)$ of $B$
in $T$ via its mgm with $T(w)$.
Now, consider the rewriting tree transition determined by the mgu $\sigma_1$, i.e. consider 
$T' \rightarrow T'_w$.
  If the branch corresponding to $B$ in
$T'_w$ does
not contain an unguarded loop, then it too must have a leaf
node $T'_w(u)$ that  unifies with
$P(k_2)$ via mgu $\sigma_2$, say, and $P(k_2)$ is exactly the clause
used to construct a node $T(vj)$ of $B$ in $T$ via its mgm with
$T(v)$.  And so on.  After some finite number $n$ of tree transitions, of ``growing" the branch corresponding to $B$ in $T'$ by taking further mgu's on its leaves,
we must come to a rewriting tree $T^*$ for $P$ and the root atom $\sigma(\mathit{head}(P(i)))$ in $D_i$
that contains an unguarded loop corresponding to $L$, and where $\sigma =
\sigma_n \circ ... \circ \sigma_1$ for the mgu's
$\sigma_1,...,\sigma_n$ involved in the tree transitions in
$D_i$. Indeed, since branches $B$ of $T$ and
$T^*$ are constructed using mgm's with
exactly the same clauses at each step, and since the mgu's
$\sigma_1,..., \sigma_n$ are all {\em most general} unifiers, we must
have that $\sigma$ is more general than $\theta$, and thus
$T^*$ is a more general version of $T$, and so
contains an unguarded loop that is a more general version of $L$.

\subsection*{Proof of Proposition~\ref{prop:fot}: Finiteness of observation subtrees}

Let $D'$ be the observation subtree of the derivation tree $D$ for a program $P$ and an atom $A$.  If $D$ is
finite, then $D'$ will necessarily be finite, so we may, without loss
of generality, suppose $D$ is infinite. 

If there exists a rewriting tree in $D$ that is unguarded, then, by
Condition 2a of Definition~\ref{def:resapp3}, the branch of $D$ on
which that tree appears will end at that tree in $D'$ and will thus be
finite. For $D'$ to be infinite, there must exist an infinite branch of
$D$ containing only guarded rewriting trees such that coinductive invariants computed in that branch never repeat. 
In fact, every infinite
branch of $D'$ must satisfy these two conditions.

Let $T$ be any guarded rewriting tree on any infinite branch of $D$.
We first note that the coinductive invariant $\kw{ci}(T)$ must be
non-empty. In addition, $T$ must itself be finite. Indeed, if $T$ were
infinite then, by the completeness of breadth-first search, an
unguarded rewriting tree would have to exist at some finite depth on
$T$'s branch of $D$. Then, by the argument of the preceding paragraph,
$T$'s branch of $D$ would have to be finite. But this is not the case.

So $T$ must be a finite, guarded rewriting tree appearing on an
infinite branch of $D$. Now, although $D$ itself is infinite,
Proposition~\ref{prop:finite} ensures that $D$'s coinductive invariant set
still contains only finitely many clause projections, so any branch of
$D$ can add only finitely many distinct elements to $D$'s coinductive invariant set. In particular, the coinductive invariants for nodes on $T$'s
infinite branch of $D$ must eventually all be equal. Moreover, since
$\kw{ci}(T) \not = \emptyset$, these coinductive invariants must eventually
all be non-empty. Thus Condition 2b of Definition~\ref{def:resapp3}
must eventually be satisfied and the branch of $D'$ corresponding to
$T$'s branch in $D$ must thus be finite. Having argued that the branch
of $D'$ corresponding to any infinite branch of $D$ is finite, we have
that $D'$ is itself finite.

\subsection*{Proof of Lemma~\ref{lem:main2}: Guardedness of observation subtree implies guardedness
of derivation tree}

The proof proceeds by induction-coinduction. We assume the observation
subtree $D'$ for $D$ is guarded and inductively examine every
branch $B'$ of $D'$. This is possible because the number of such
branches and their lengths are all finite by Proposition~\ref{prop:fot}. For any
such $B'$, either no parent of any leaf in the last coinductive tree
of $B'$ can be resolved with any clause of $P$, or $B'$ was terminated
by Condition 2b of Definition~\ref{def:resapp3}. In the former case, the entire
branch $B'$ will also appear in $D$, and each rewriting tree on the
corresponding branch $B$ of $D$ will be guarded. In the latter case,
we can proceed coinductively.

If $B'$ was terminated because it contains a guarded transition $T
\Longrightarrow T'$ for $T = D(w)$ and $T' = D(wv)$, then both $T$
and $T'$ were formed by resolving with some clause $P(k)$. In this
case, we apply the following coinductive argument. \emph{Coinductive
  Hypothesis (CH): The process of resolving with clause $P(k)$ to
  produce a new guarded rewriting tree whose coinductive invariant has
  first component $P(k)$ can be repeated infinitely many times in
  transition sequences originating from $T$.} 
  By computing that CH is again satisfied for $T'$, we can make the
  following {\em Coinductive Conclusion (CC): For any tree $T$ in any
    branch $B$ containing $B'$, the process of resolving with clause
    $P(k)$ to produce a new guarded rewriting tree whose coinductive invariant has first component $P(k)$ can be repeated infinitely many
    times in transition sequences originating from $T$.}
So each of the rewriting trees in $B$ must be guarded.

Unfortunately, CC does not guarantee that no unguarded loop can
possibly occur in $D$ by resolving with other clauses in the
sequence of transitions from $T'$ that occur in $B$ but not in
$B'$. But if it is possible to compute a sequence of rewriting tree
transitions in $D$ from $T'$ involving mgus $\theta_1, \ldots ,
\theta_n$ computed by resolving with clauses $P(k_1), \ldots , P(k_n)$
that lead to an unguarded rewriting tree in $B$, then, by completeness of the breadth-first construction of the derivation tree $D$,
 there must be a
a rewriting tree $T^*$ occurring in the sequence of rewriting trees in
$B'$ from $T$ to $T'$ that leads to a sequence of rewriting tree
transitions in another branch $B''$ of $D$ involving exactly the
same sequence $P(k_1), \ldots , P(k_n)$ of clauses and mgus
$\theta'_1, \ldots , \theta'_n$ such that, for each $i\in
\{1,...,n\}$, 
$\theta'_i = \sigma_i \circ \theta_i$ for some $\sigma_i$.  This holds
because rewriting tree transitions only lead to further instantiations
of variables, and the rewriting tree $T^*$ appears earlier
on $B$ than $T'$ does, and hence is more general. But then an unguarded loop induced by the mgus
$\theta'_1, \ldots , \theta'_n$ obtained by resolving with $P(k_1),
\ldots , P(k_n)$ will be found in one of the branches of $D'$ to
which $T^*$ leads.

By inducting on all branches of the observation subtree $D'$ of
$D$, and coinductively terminating each, we conclude that if all
branches of $D'$ are terminated by the above coinductive argument
with no unguarded rewriting tree being found, then no unguarded loop
can exist in any of the rewriting trees of $D$.

\subsection*{Proof of Lemma~\ref{lem:GC3-t}: GC3 is decidable}

To decide guardedness of logic programs, we must let $P$ be given and construct a set of derivation trees, one derivation tree for each clause head of $P$,
i.e. every such $D_i$ is a derivation tree for $P$ and  $\mathit{head}(P(i))$.
Moreover, we build these trees only until we construct the observation subtree $D'_i$ for each $D_i$.
We next check  whether or not each observation subtree $D'_i$ 
 is guarded. That is, we must
check whether or not every rewriting tree in $D'_i$ is
guarded and  whether or not condition 2.a of Definition~\ref{def:resapp3} was used to construct $D_i'$. Since guardedness of rewriting trees is decidable by
Proposition~\ref{prop:ct-guard-dec}, and there are only finitely many rewriting trees in
any observation tree $D'_i$, guardedness of all observation subtrees $D_i'$ for this program is decidable.
Since, by Lemma~\ref{lem:main2}, guardedness of observation subtrees implies guardedness of derivation trees,
guardedness of $P$ is also decidable.


\hfill

\section*{Appendix B. Implementation of Observational Productivity Checks}

\subsection*{Algorithmic overview of observational productivity checking }
Definitions of contraction ordering, guarded rewriting trees, and
observation subtrees translate naturally into algorithmic
forms that give rise to the implementation of our observational
productivity checks~\cite{Martin}. Below we give a high-level
pseudocode representation of the formal definitions of this paper.

\begin{algorithm}                      
\caption{Observational productivity check for a logic program}          
\label{alg1}                           
\begin{algorithmic}                    
\Require $P$ -- a logic program over signature $\Sigma$
\Require $LC$ -- an empty list
	\State $n = $ number of clauses in $P$
	 \For{$i = 0, \ldots , n $}
					\If {observation subtree $D'$ of the derivation tree $D$ for $P$ and $\mathit{head}(P(i))$ is unguarded}
                        \State { $P(i)$ is not guarded. }
                    \Else       
                      \State { $P(i)$ is guarded. } 
				      \If {$D'$ contains transition $D(v) \Longrightarrow D(w)$ with coinductive invariant $c$}
							
						   \State $LC$ := $append (LC, c)$
						   \EndIf
				\EndIf				     		
	   \EndFor
				\If {all $P(i)$ are guarded}
                     \State $Result1$ := ``$P$ is guarded"
				\Else 
				     \State $Result1$ := ``$P$ is not guarded" 
				\EndIf
				\If {$LC$ is not empty}
                    \State $Result2$ := ``$P$ is existentially live with $LC$"
				\Else 
				    \State $Result2$ := ``$P$ has finite derivations only"
				\EndIf
    \State \Return ($Result1, Result2$)
\end{algorithmic}
\end{algorithm}

Algorithm~\ref{alg1} below captures the essence of our check that
$GC3(P)$ holds for a logic program $P$. It depends on the definition
of the observation subtree (Definition~\ref{def:resapp3}), which
in turn depends on two conditions:
\begin{itemize}
\item finiteness of observation subtrees, as proven in
  Proposition~\ref{prop:fot}, and
\item guardedness of every rewriting tree in a program's observation subtree.
\end{itemize}
These two conditions ensure termination of Algorithm~\ref{alg1}, as
expressed formally in Lemma~\ref{lem:GC3-t}.  In the main body of this
paper we have written $GC2(T)$ to indicate that a rewriting tree $T$
is guarded. A pseudocode description of our check that $GC2(T)$ holds
for a rewriting tree $T$ is given in Algorithm~\ref{alg2}.

 \begin{algorithm}                      
\caption{Guardedness check in  a rewriting tree}          
\label{alg2}                           
\begin{algorithmic}                    
\Require $T$ --  the rewriting tree for a logic program $P$ and an atom $A$
    \For{$i = 0, \ldots , depth(T) $}
        \For{nodes $w_1, \ldots w_m$ at depth $i$}
            \If {a node $w_j$ forms a loop  with some node $v$ above it}
                \If {$loop(T,v,w_j)$ is not guarded}
                    \State \Return ``$T$ is not guarded"
                \EndIf
            \EndIf	     		
	   \EndFor
	\EndFor
    \State \Return ``$T$ is guarded"
\end{algorithmic}
\end{algorithm}

Termination of Algorithm~\ref{alg2} depends crucially on
Proposition~\ref{prop:ct-guard-dec}, i.e., on the fact that it is
impossible to construct an infinite rewriting tree without finding
unguarded loops. Algorithm~\ref{alg2} in turn relies on an algorithmic
check that two terms are related via a contraction ordering, but we
omit specifying this in pseudocode since it is entirely
straightforward.

\subsection*{Implementation}

Our observational productivity checker is implemented in Go
(golang.org) as a command line program and is part of the general
implementation of structural resolution and coalgebraic logic
programming (CoALP)~\cite{Martin}.  Go was chosen as implementation
language because it provides easy primitives for parallelization,
which has been explored to optimize proof search~\cite{KSH14}. To compile and install the productivity checker follow the
instructions in the README file supplied in the program distribution
available at \cite{Martin}.

CoALP can be used not only to check the productivity of logic
programs, but to make queries to guarded such programs as well. The
checker takes Prolog-style programs saved in text files as input. The
format of programs corresponds exactly to that of Prolog. For example,
program $P_6$ is represented as

\begin{verbatim}
from(X, scons(X, Y)) :- from(s(X), Y).
\end{verbatim}

\noindent
Unlike Prolog, our checker does not support built-in predicates or
arithmetic functions.

To check a logic program for observational productivity, the path to
the program file has to be given as the first parameter:

\begin{verbatim}
guardcheck somefile.logic 
\end{verbatim}

\noindent
The above command initialises the $GC3(P)$ check for a given logic
program $P$ in $\mathtt{somefile.logic}$, and, as $GC3$ involves
computations of coinductive invariants, it simultaneously uses them to
detect existential liveness, as detailed in Algorithm~\ref{alg1}.
Many example files and tests as well as the programs used in this
paper can be found in the directory named ``examples" in
\cite{Martin}.

The output for the observationally productive program $P_6$ is:
\begin{verbatim}
Program is guarded.
Program is existentially live with coinductive invariants:
in clause 0 of "from": [{0 | scons(v3,v5) | [1]}]
\end{verbatim}

Note that the first  \texttt{0} in \texttt{ [{0 | scons(v3,v5) | [1]}]} points to the clause $P_6(0)$, and would suggest that the predicate
$\mathtt{from}$
 in the head of this clause is a good candidate to be given coinductive semantics and hence coinductive typing.
We believe that this general information can be used by CoLP or CoALP to determine typing for coinductive predicates in their programs.

The output for the unguarded program $P_7$ is:
\begin{verbatim}
Program is not guarded.
Goal q(s(v34),s(v34),s(v42),v36) results in unguarded loop 
in path [(p:0), (q:0), (p:0)].
\end{verbatim}

\subsection*{A more complex example}

In this section, we consider a more challenging example of \emph{Sieve
  of Eratosthenes}, known for its difficulty in the literature on
coinductive definitions~\cite{GuptaBMSM07}. The following program
$P_9$ is an observationally productive reformulation of the original
\emph{Sieve of Eratothsenes} program from~\cite{GuptaBMSM07}:

\vspace*{0.05in}

\noindent
$0.\; \mathtt{prime(X)} \gets \mathtt{inflist(I), sieve(I,L), member(X,L)}$\\
$1.\; \mathtt{sieve(cons(H,T),cons(H,R))} \gets \mathtt{filter(H,T,F), sieve(F,R)}$\\
$2. \;\mathtt{filter(H,cons(K,T),cons(K,T1))} \gets \mathtt{mod(X,K,H), less(0,X),filter(H,T,T1)}$\\
$3. \;\mathtt{filter(H,cons(K,T),T1)} \gets \mathtt{mod(0,K,H), filter(H,T,T1)}$\\
$4. \;\mathtt{int(X,cons(X,Y))} \gets \mathtt{int(s(X),Y,Z1)}$\\
$5. \;\mathtt{inflist(I)} \gets \mathtt{int(s(s(0)),I)}$\\
$6. \;\mathtt{member(X,cons(X,L))} \gets$\\
$7. \;\mathtt{member(X,cons(Y,L))} \gets \mathtt{member(X,L)}$\\
$8. \;\mathtt{less(0,s(X))} \gets $ \\
$9. \;\mathtt{less(s(X),s(Y))} \gets \mathtt{less(X,Y)}$\\

\noindent
The original program~\cite{GuptaBMSM07} does not use Prolog-style list notations and
uses a structural representation of numbers, which we avoid.  We also assume a suitable implementation of the modulo operator as the
predicate $\mathtt{mod}$ above.

If we run the observational productivity check on this program, we obtain the following output:
\begin{verbatim}
Program is guarded.
Program is existentially live with coinductive invariants:
in clause 0 of "filter": [{0 | cons(v16,v17) | [1]} {0 | cons(v16,v18) | [2]}]
in clause 1 of "filter": [{1 | cons(v24,v25) | [1]}]
in clause 0 of "sieve": [{0 | cons(v10,v11) | [0]} {0 | cons(v10,v12) | [1]}]
in clause 1 of "member": [{1 | cons(v37,v38) | [1]}]
in clause 1 of "less": [{1 | s(v40) | [0]} {1 | s(v41) | [1]}]
\end{verbatim}

Above, four predicates have been identified as potentially having a coinductive semantics: 
\texttt{filter}, \texttt{sieve}, \texttt{member} and \texttt{less}.
Generally, most inductive definitions admit coinductive interpretation, and predicates that we intuitively consider as inductive
may be identified as potentially coinductive. This situation was analysed in Example~\ref{ex:ordinal}.
Among the four predicates, \texttt{sieve}, that admits only coinductive interpretation, was identified.

We note that the original formulation of~\cite{GuptaBMSM07} is not
observationally productive, since it does not possess universal
observability property.  If we run our checker on the formulation
in~\cite{GuptaBMSM07}, failure of observational productivity is
detected and reported as follows:
\begin{verbatim}
Program is not guarded.
Goal comember(v136,v140) results in unguarded loop
in path [(primes:0), (comember:0), (comember:0)].
\end{verbatim}
\noindent
As indicated by the checker output, the reason is that the following
definition of {\em comember} used in~\cite{GuptaBMSM07} is not
universally observable and hence is not guarded:

\vspace*{0.05in}

\noindent
$0.\; \mathtt{comember(X,L)} \gets \mathtt{drop(X,L,L1),
  comember(X,L1)}$\\ $1.\; \mathtt{drop(H,cons(H,T),T)} \gets
$\\ $2.\; \mathtt{drop(H,cons(H,T),T)} \gets \mathtt{drop(H,T,T1)}$

\vspace*{0.05in}

\noindent
Indeed, the definition of $\mathtt{comember}$ in Clause $0$ above is
not guarded by any constructors.

In our reformulation as program $P_9$ above, we use a guarded
definition of $\mathtt{member}$ instead of the definitions of
$\mathtt{comember}$ and $\mathtt{drop}$ used in~\cite{GuptaBMSM07}.
The definition of $\mathtt{member}$ is guarded by the constructor
$\mathtt{cons}$ in Clause $7$. Thus, in the case of the Sieve of
Eratothsenes, the transition from an unproductive to a productive
coinductive definition was a simple matter of applying a program
transformation that clearly preserves the intended coinductive meaning
of the coinductive definition of $\mathtt{seive}$ in Clause $1$.

\end{document}